\documentclass[twocolumn,showpacs,preprintnumbers,amsmath,amssymb,superscriptaddress]{revtex4}

\usepackage{graphicx,subfigure}
\usepackage{dcolumn}
\usepackage{bm}
\usepackage{psfrag}
\usepackage{multirow}

\def\be{\begin{equation}}       \def\ee{\end{equation}}
\def\bea{\begin{eqnarray}}      \def\eea{\end{eqnarray}}

\begin{document}
\title{1D-FFLO state in the absence of time-reversal symmetry breaking}
\author{Trinanjan Datta}
\affiliation{Department of Chemistry and Physics, Augusta State
University, Augusta, GA 30904}\email{tdatta@aug.edu}
\date{\today}
\begin{abstract}\label{abstract}
A novel route to a one-dimensional Fulde-Ferrell-Larkin-Ovchinnikov (1D-FFLO) state in the absence of broken time-reversal symmetry is proposed in this paper. At present such a state may be encouraged in a clean AlAs quantum wire. Using the AlAs quantum wire as an example it is shown using bosonization and the renormalization group approach that the 1D-FFLO state can arise due to a combination of Coulomb interactions and the unique bandstructure arrangement of the AlAs quantum wire. The present theoretical proposal is very general and is applicable to other systems with similar fermionic interaction terms.
\end{abstract}

\maketitle

A superconducting state with a finite pairing momentum had been long proposed by Fulde and Ferrell \cite{FuldeFerrell}, and Larkin
and Ovchinnikov \cite{LarkinOvchinnikov}. Presently this state is known as the Fulde-Ferrell-Larkin-Ovchinnikov (FFLO) (or LOFF) state. The FFLO state is widely realized from condensed matter systems to high energy physics \cite{casalbuoni:263}. In a FFLO state Cooper pairs with a nonzero center-of-mass momentum are encouraged in the presence of an external magnetic field. The central physical ingredient of the originally studied FFLO state is based on the idea that Cooper pairs in a spin-singlet superconductor are composed of fermions with opposite spins. In the presence of a magnetic field these electrons couple to the external applied field through the Zeeman coupling. This in turn tends to polarize the electrons along the direction of the magnetic field and allows the system to gain polarization energy. However, the pairing of opposite spins is favorable for condensation energy. Due to these competing tendencies the superconducting state undergoes a transition to the FFLO state with a finite pairing momentum and eventually enters the normal state as the magnetic field is further increased \cite{FuldeFerrell,LarkinOvchinnikov}. 

There have been several theoretical proposals that such a state may be achievable
in heavy fermion systems and organic and cuprate superconductors \cite{gloos:501,tachiki:369,modler:1292,gegenwart:307,brien:1584,singleton:641,singh:187004,tanatar:134503,krawiec:134519}, as well as 
in cold atomic gases \cite{liu:047002}, 1D Kondo lattice \cite{oron:1342}, and even nuclear and quark matter \cite{casalbuoni:263}.
The field has been energized by recent experimental evidence of the FFLO state in 
the heavy fermion superconductor CeCoIn$_5$ \cite{cui:214514,radovan:51,bianchi:187004}.
Recent theoretical progress has indicated that the state may be
achievable through applied current rather than an applied field \cite{doh:257001}.
The original prediction as well as the more recent ones rely on introducing a magnetic field 
(whether externally or as an effective internal mean field \cite{machida:122}), or using some
other perturbation in order to create the unequal Fermi surfaces
necessary to generate finite-momentum pairing. 

In this paper, a novel route to 
a 1D-FFLO state is proposed
\emph{in the absence of a magnetic field or any other external perturbation} with the AlAs system as an example. This system 
was recently fabricated and investigated by Moser {\em et al.} \cite{moser:052101,moser:193307}.
In such a system it should be possible to induce an FFLO state due to the interplay of Coulomb interaction effects with the specific bandstructure of the AlAs quantum wire, which includes a unique inter-valley umklapp scattering process that is present {\em at all densities}. 
\begin{figure}[h]
\centering 
  {\label{fig:bandstructure}{\psfrag{k}{k}\psfrag{E}{E}\psfrag{umk}{$\frac{k_{U}}{2}$}\psfrag{A}{A}\psfrag{B}{B}
  \includegraphics[width=.95\columnwidth]{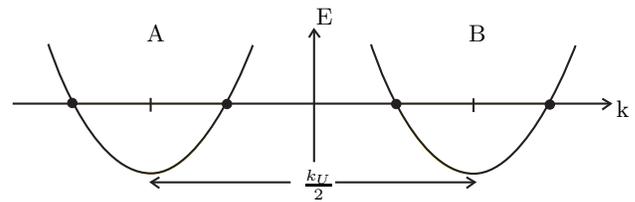}}}
\caption{Schematic of the 1D bandstructure in the AlAs quantum wire of Ref.~\onlinecite{moser:052101}.
The bandstructure has two degenerate subbands at the lowest densities, leading to four Fermi points. In addition, the band minima are separated by half an umklapp vector, $k_{U}/2$, giving rise to umklapp interactions which are present at all fillings. The effective dispersion can be linearized so that the Fermi points, represented by black dots on the figure, are at $k^{A\pm}_{F}=
-\frac{k_{U}}{4}\pm k^{o}_{F}$ and $k^{B\pm}_{F}= \frac{k_{U}}{4}\pm
k^{o}_{F}$ where $k_U$ is the umklapp vector, and $k_{F}^{o}$ is the
magnitude of the Fermi wavevector measured from the bottom of each
band. 
\label{fig:bandstructure}}
\end{figure}

A schematic of the AlAs quantum wire bandstructure is shown in Fig.~\ref{fig:bandstructure} \cite{moser:052101,moser:193307}. 
While bulk AlAs has three degenerate bands, 
the growth direction and cleavage planes chosen in fabricating the wire leave two degenerate valleys at low energy.
In the effective one-dimensional bandstructure, this leads to two degenerate nonoverlapping bands separated by 
half an umklapp vector, as shown in Fig.~\ref{fig:bandstructure}. Because the momenta of the two band minima are related by half an umklapp vector, there is a class of umklapp excitations unique to this bandstructure which exists {\em at all densities}. 
Because there are four Fermi points, the bandstructure supports several fermionic scattering processes.
We retain only those scattering terms which are relevant in a renormalization group sense.

\begin{figure}[h]
 {\psfrag{E}{E}\psfrag{k}{k}\label{fig:umklappcooper}\includegraphics[width=3.5in]{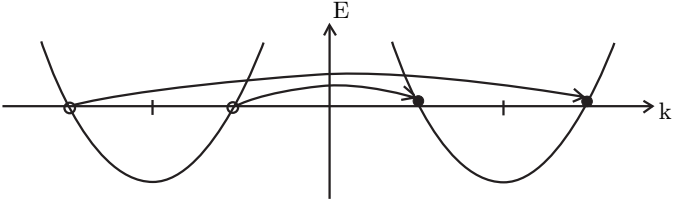}}
\caption{Representative inter-valley umklapp scattering process which is
 responsible for encouraging the 1D-FFLO state.
In the process which is shown, a pair of left and right movers from the A band 
scatters into a pair of left and right movers in the B band by absorbing an
umklapp vector.  
This is an umklapp process which is \emph{present
at all densities}.
\label{fig:umklappcooper}}\end{figure}

In a quantum wire electrons are quantum mechanically confined to move along
one direction with their motion in the remaining transverse
directions confined via a potential, $V_{conf}(\vec{r_{\perp}})$, 
where $\vec{r}_{\perp}=(y,z)$ denotes the transverse coordinates of
quantization. Electron-electron interactions within the wire are
described by $U(\vec{r}$) which is purely repulsive. The Hamiltonian
is a sum of two independent terms in the transverse and longitudinal
directions with the result that the wavefunction (and therefore the
correlation functions) can be decomposed as a product of $
\phi(\vec{r}_{\perp})$ and $\psi_{s}(x)$ where
$\phi(\vec{r}_{\perp})$ is the orthogonal wavefunction of transverse
quantization of the two degenerate bands ($X$ and $Y$ valleys) and
$\psi_{s}(x)$ the longitudinal part. In order to describe the
physics along the longitudinal direction we now promote the
wavefunction, $\psi_{s}(x)$, to the level of a field operator (for a
field theoretic description) responsible for creating and
annihilating the electrons taking part in the various scattering
processes. With this in mind the second quantized Hamiltonian
suitable for 
the purposes of our study is
\begin{eqnarray}\label{eq:hamiltonian}
H = \sum_{s} \int d^{3}r \Psi^{\dag}_{s}(\vec{r})
\left(-\frac{1}{2m}\vec{\nabla}^{2}_{r} -\mu +
V_{conf}(\vec{r}_{\perp})
\right )\Psi_{s}(\vec{r}) \nonumber \\
+ \frac{1}{2}\sum_{s,s^{'}} \int d^{3}r
d^{3}r^{'}U(\vec{r}-\vec{r^{'}})
\Psi^{\dag}_{s}(\vec{r})\Psi^{\dag}_{s^{'}}(\vec{r}^{'})\Psi_{s^{'}}(\vec{r}^{'})\Psi_{s}(\vec{r})
\nonumber \\
\end{eqnarray}
where $\Psi_{s}(\vec{r})= \phi(\vec{r}_{\perp})\psi_{s}(x) $ is now
the field operator for an electron species of spin
$s=\{\uparrow,\downarrow$\}, and $\mu$ is the chemical potential in
the leads. 
Because the low energy, long wavelength excitations occur in 
the vicinity of the Fermi points (see
Fig.~\ref{fig:umklappcooper}) a further decomposition is possible
with $\Psi_{s}(\vec{r})= \phi(\vec{r}_{\perp})(\psi_{As}(x) +
\psi_{Bs}(x))$. The coordinate $x$ is in the long direction of the
wire. The longitudinal part of the field can be naturally expanded
in terms of the right- and left- moving excitations, $R_{ns}(x)$ and
${L}_{ns}(x)$, respectively, residing around the Fermi points of the
two bands (indicated by the black dots in
Fig.~\ref{fig:bandstructure}) with $\psi_{As}(x)=
R_{As}(x)e^{ik^{A+}_{F}x} + L_{As}(x)e^{ik^{A-}_{F}x}$ and
$\psi_{Bs}(x)= R_{Bs}(x)e^{ik^{B+}_{F}x} +
L_{Bs}(x)e^{ik^{B-}_{F}x}$. The band index is $n=A,B$ and the Fermi
momenta are defined by $ k^{A\pm}_{F}=-\frac{k_{U}}{4}\pm k^{o}_{F}$
and $k^{B\pm}_{F}= \frac{k_{U}}{4}\pm k^{o}_{F}$ where $k_U$ is the
umklapp vector, and $k_{F}^{o}$ is the magnitude of the Fermi
wavevector measured from the bottom of each band, as shown in
Fig.~\ref{fig:bandstructure}.

The low energy properties of the interacting 1D electron gas can now be
conveniently described within the framework of the bosonization
technique \cite{Giamarchi}. 
Within this approach, the fermionic field operators can be
written in terms of bosonic fields
$\phi_{n\nu}=(\phi_{n\uparrow}\pm\phi_{n\downarrow})/ \sqrt{2}$ and
$\theta_{n\nu}=(\theta_{n\uparrow}\pm\theta_{n\downarrow})/
\sqrt{2}$, where $\nu = \rho,\sigma$ (the charge and spin modes)
correspond to the $\pm$ combination, $s=\{\uparrow,\downarrow$\} is
the spin index and $n=A,B$ is the band index. The bosonic fields
satisfy the commutation relation
[$\phi_{n\nu}(x),\partial_{x^{'}}\theta_{n^{'}\nu^{'}}(x^{'})$]=i$\pi\delta(x-x^{'})\delta_{\nu,\nu^{'}}\delta_{n,n^{'}}$
with $\hbar$ set equal to one. We then have
\begin{equation}
{R}_{ns}(x) = \frac{\eta_{Rns}}{\sqrt{2\pi\alpha}}e^{i\sqrt{\pi
\over 2} [\theta_{n\rho}(x)- \phi_{n\rho}(x) +s(\theta_{n\sigma}(x)
- \phi_{n\sigma}(x))]}
\end{equation}and,
\begin{equation}
{L}_{ns}(x) = \frac{\eta_{Lns}}{\sqrt{2\pi\alpha}}e^{i\sqrt{\pi
\over 2} [\theta_{n\rho}(x)+ \phi_{n\rho}(x) +
s(\theta_{n\sigma}(x)+ \phi_{n\sigma}(x))]}
\end{equation}
where $\alpha$ is the short distance cutoff, 
and 
$\eta_{Rns}$ and
$\eta_{Lns}$ are the Klein factors for the right- and left- moving
fields of band $n$ with species of spin $s$. They are required to
preserve the anti-commutation relations of the fermionic fields. The
convenient field variables for the Hamiltonian in our problem will
be a linear combination of the boson fields constructed out of the
two bands. We define the transformation to a symmetric and an
anti-symmetric basis as $\phi^{\pm}_{\nu}
=\frac{1}{\sqrt{2}}(\phi_{A\nu} \pm \phi_{B\nu})$ and
$\theta^{\pm}_{\nu} = \frac{1}{\sqrt{2}}(\theta_{A\nu} \pm
\theta_{B\nu})$. 

Upon bosonization and subsequent transformation the
parts of the Hamiltonian corresponding to 
kinetic energy lead to 
harmonic terms in the symmetric and the anti-symmetric bosonic
fields $(\phi^{\pm}_{\nu}$ and $\theta^{\pm}_{\nu})$. 
The
intraband interactions and the interband interactions, however,
generate both cosine interaction terms and harmonic terms \cite{umklappnote}. 
The Hamiltonian, $H$, can then be written in the following canonical form.
\begin{equation}\label{eq:bosonizedhamiltonian}
H=\frac{1}{2}\sum_{\nu=\rho,\sigma}\int dR
\bigg[v^{\pm}_{\nu}K^{\pm}_{\nu}(\partial_{R}\theta^{\pm}_{\nu})^{2}
              + \frac{v^{\pm}_{\nu}}{K^{\pm}_{\nu}}(\partial_{R}\phi^{\pm}_{\nu})^{2}\bigg]+H_{int}~.
\end{equation}
Here,  $R=(x+x^{'})/2$  is the center-of-mass
coordinate of two electrons and $a= x -x^{'}$ their relative
coordinate in the long direction of the quantum wire. In the quadratic part
of the Hamiltonian 
the bare symmetric and anti-symmetric Luttinger parameters,
$K^{\pm}_{\nu}$ can be expressed in terms of the original Luttinger parameters $K_{\nu}$.
The symmetric and anti-symmetric velocities $v^{\pm}_{\nu}$ can also be expressed in terms of the original
velocities $v_{\nu}$.
The terms responsible for interactions in the bosonic fields are included  in 
$H_{int}$.
There are eleven such interaction terms in the bosonized form.
Weak coupling renormalization group (RG) (to second order in the interaction coupling constants)
is then used to determine the instabilities of the AlAs quantum wire \cite{balents-1996,schulz-1996,lin-1998}.
The corresponding coupled nonlinear differential equations are then solved, using
appropriate physical inputs as initial conditions for the RG flow,
as discussed below.

The 
interaction terms which ultimately
determine the final phases are 
\begin{eqnarray}\label{interactions}
&&\frac{t_{a}}{2\pi^{2}\alpha^{2}}\int dR \cos[\sqrt{4\pi}\phi^{-}_{\rho}]\cos[\sqrt{4\pi}\phi^{+}_{\sigma}] \nonumber \\
&&\frac{t_{b}}{2\pi^{2}\alpha^{2}}\int dR \cos[\sqrt{4\pi}\phi^{-}_{\rho}]\cos[\sqrt{4\pi}\phi^{-}_{\sigma}] \nonumber \\
&&\frac{t_{c}}{2\pi^{2}\alpha^{2}}\int dR  \cos[\sqrt{4\pi}\theta^{-}_{\rho}]\cos[\sqrt{4\pi}\phi^{+}_{\sigma}] \nonumber \\
&&\frac{t_{d}}{2\pi^{2}\alpha^{2}}\int dR \cos[\sqrt{4\pi}\theta^{-}_{\rho}]\cos[\sqrt{4\pi}\phi^{-}_{\sigma}]~.
 \end{eqnarray}
The initial values of the coupling constants are set by 
\begin{eqnarray}
\label{initconditions}
&&t_{a}=2\int da M(a)\cos[2k^{o}_{F}a]  \nonumber \\
&&t_{b}= 2\int da M(a)\bigg(\cos[2k^{o}_{F}a]-\cos\bigg[\frac{k_{U}a}{2}\bigg]\bigg) \nonumber \\
&&t_{c}= 2\int da M(a)\cos[2k^{o}_{F}a]\cos\bigg[\frac{k_{U}a}{2}\bigg] 
 \nonumber \\
&&t_{d}= 2\int M(a)\cos\bigg[\frac{k_{U}a}{2}\bigg] 
\end{eqnarray} 
We use a screened Coulomb potential for the
interaction kernel, 
$M(a) =\frac{e^{2}}{4\pi\epsilon}\frac{e^{-(a/d)}}{\sqrt{w^{2}+
a^{2}}}$, 
where
$e$ is the electronic charge, 
$\epsilon = 10.9 \epsilon_{o}$ is the dielectric constant of AlAs \cite{dielectric}, and
$\epsilon_{o}$ is the permittivity of free space.
The width of the quantum wire $w$ provides a short distance cutoff
whereas the distance to the back-gate $d$ provides a long distance cutoff. 

\begin{figure}[htb]
{\centering 
 \subfigure[~$n=10^5 cm^{-1}$\label{nE5percm}]{\includegraphics[width=.95\columnwidth]{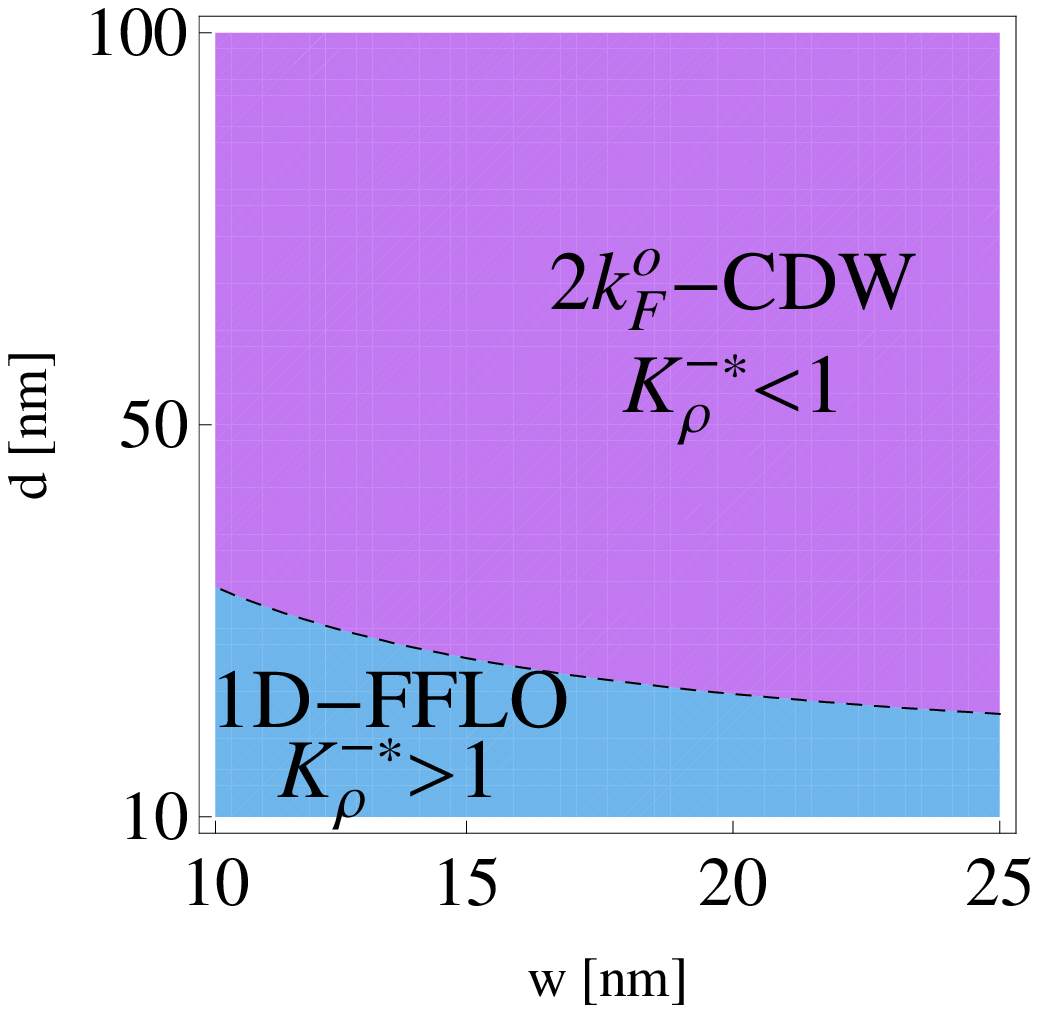}}
 \subfigure[~$n=5 \times 10^4 cm^{-1}$\label{n5E4percm}]{\includegraphics[width=.95\columnwidth]{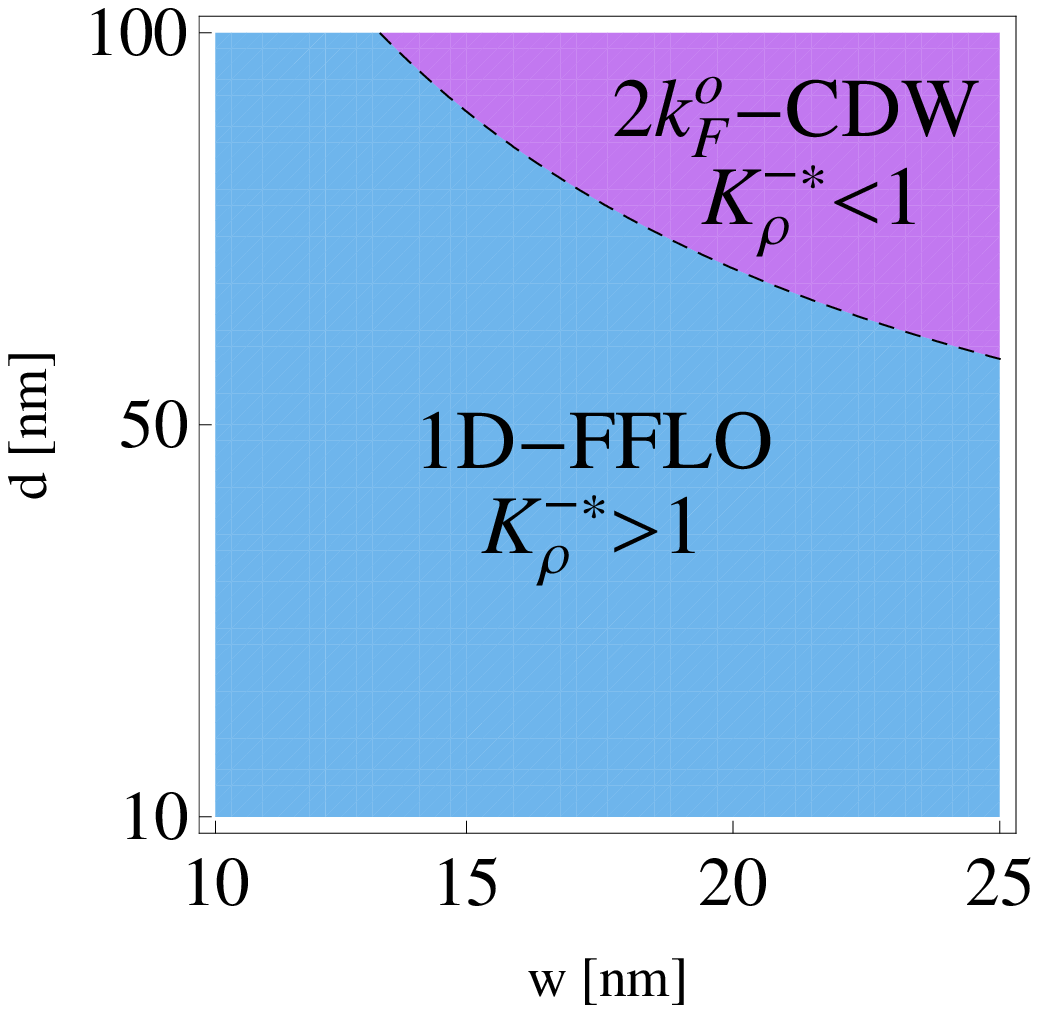}}
}
\caption{Variation of $K^{-}_{\rho}$ with distance to the back gate  $(d)$ and width of the wire $(w)$,
using $L= 1\mu m$,  
$\epsilon = 10.9 \epsilon_o$, and $m^* = 0.33 m_e$.
\label{fig:Krhominusvariationcontourplot}}
\end{figure}

Using input parameters appropriate to the wire, and the expression
for the coupling constants, Eq.~\ref{initconditions}, we can make an
estimate for the initial conditions of the RG flow, {\em i.e.} the bare values of the coupling constants $t_i$.
The AlAs quantum wire of Moser \emph{et al.}\cite{moser:193307,moser:052101}  fabricated using standard design parameters has a transverse size of $w \approx 15 nm$ separated from the metallic gate
by a distance $d \approx 300 nm$. The wire length is $L \approx 1 \mu m$,
so that  $w/d \approx$ 0.05 and  $L/d$ = 10/3. 
The parameters used to estimate $k^{o}_{F}$ for the AlAs bandstructure
are as follows: the density of the electrons in the quantum wire $\approx 10^{6}cm^{-1}$ and the
effective mass of the electron $m^{*}= 0.33m_{e}$ along the long
direction where $m_{e}$ is the bare mass of the electron \cite{moser:193307}. 
Experimental evidence \cite{Gunawan,Shkolnikov} suggests that spin rotational invariance is not broken in the AlAs
quantum wells in the absence of a magnetic field ({\em i.e.} spin-orbit coupling is negligible.) Furthermore, since there is no external magnetic field present spin rotational invariance is not broken and we set the initial spin Luttinger parameters $K^{\pm}_{\sigma}$ equal to one.

Given these inputs for the bare parameters, we 
allow the running coupling constants to flow according to the RG equations with initial conditions for the bare Luttinger parameters $K^{-}_{\rho}<1$ or $K^{-}_{\rho}>1$ together with initial values for $K^{\pm}_{\sigma}$ set equal to one. When one or more coupling constants grows to be of $\mathcal{O}(1)$, 
we stop the flow.  At that point, the fields for the corresponding
divergent couplings develop a gap to excitations, and acquire
a definite expectation value.  This expectation value is then 
inserted into the correlation functions in order to determine the
type of phase this instability represents.

Using this standard analysis, with initial conditions appropriate to the
AlAs quantum wire as described above, we find that for the wire fabricated by Moser {\em et al.} 
the coupling constants which diverge are $t_a$ and $t_b$, leading to a state with divergent $2k^{o}_{F}$-charge-density wave $(CDW)$ correlations. This  arises because the relative charge channel flows to an effectively repulsive state, with effective Luttinger parameter $K_{\rho}^{-*} < 1$.  

Although the bare interactions are (screened) repulsive Coulomb interactions,
it is possible for the interactions of the relative charge channel
to flow to an effective attractive regime, {\em i.e.} $K_{\rho}^{-*} > 1$.
When this happens, the coupling constants which diverge are 
$t_{c}$ and $t_{d}$, 
while $t_a$ and $t_b$ remain small.
From the interaction terms we can then deduce that the dual antisymmetric charge field
$\theta_{\rho}^{-}$ gets gapped together with the spin fields
$\phi^{\pm}_{\sigma}$. These gapped fields lead to a state with divergent intraband singlet superconductivity  correlations with a finite pairing momentum - {\bf the 1D-FFLO state}. The pairing mechanism in this case is shown in Fig.~\ref{fig:umklappcooper}.
The diverging interactions allow left and right movers from band ``A'' to scatter into a left and right mover in band ``B'' by absorbing an umklapp vector. That is, pairs form within each band in order to take advantage
of the lowering of energy possible with umklapp scattering from
one band to the other. This is analogous to the spin gap proximity effect mechanism
described in Ref.~\onlinecite{spingap}. Because each pair resides within a single
valley, the pairs each have a net momentum,  $\pm k_U/2$,
leading to a 1D FFLO state \cite{cntube, ando:2857}.

Phase diagrams showing this transition from $2k^{o}_{F}$-CDW
to the FFLO state are shown in Fig.~\ref{fig:Krhominusvariationcontourplot}.
Our calculations indicate that the previously fabricated AlAs QWR of Refs.~\onlinecite{moser:052101,moser:193307}
would require extremely low densities 
({\em i.e.} $n \approx 10^4 cm^{-1}$) in order to achieve the FFLO state.
Such low densities would not only make the behavior of the wire more susceptible
to disorder, but it would also require a longer wire in order to achieve a reasonable
total number of electrons in the wire. 
By using an alternate structure with a metallic side gate, it may be possible to
make the long distance screening length as small as $d=10-30$ nm \cite{Mattsetup}.
In this case, our calculations indicate that the FFLO state could be achievable
for more reasonable densities, on the order of $n \approx 10^5 cm^{-1}$.

In conclusion, we have proposed a novel route to an FFLO state in 1D. The state should be achievable in AlAs QWR's for which the bandstructure leads to pairing generated by  umklapp pair scattering which is present at all densities.  This particular FFLO state is intrinsic to the wire, and does not require an external field or perturbation in order to induce the state. The theoretical proposal is general. It is applicable to other systems with a similar bandstructure where analogous fermionic interaction terms are allowed.

The author wishes to acknowledge Erica W. Carlson for suggesting the problem and additionally wishes to thank Matthew Grayson and Joel Moser for numerous insightful and useful discussions about the aluminum arsenide quantum wire system. 
\bibliography{1DFFLO}

\begin{thebibliography}{33}
\bibitem[{umk()}]{umklappnote}
\bibinfo{note}{While bosonizing I have used the fact that the exponential $e^{i
  k_{U}x}=1$. This is true because the electrons in the QWR move in an
  underlying lattice and their spatial coordinate $x=ma_{x}$, where $a_{x}$ is
  the lattice spacing in the long direction of the wire and $m$ is an integer.
  When multiplied with the umklapp vector, $k_{U}=2\pi/a_{x}$, and
  exponentiated the result is $\exp[-2m\pi]=1$.}
\bibitem[{cnt()}]{cntube}
\bibinfo{note}{The 1D-FFLO correlation will NOT arise in carbon nanotubes since
  the umklapp scattering process is not allowed in the carbon nanotube
  bandstructure. This is due to the absence of backscattering processes
  \cite{ando:2857}.}
\end{thebibliography}
\end{document}